Enabling Machine Learning Algorithms for Credit Scoring - Explainable Artificial Intelligence (XAI) methods for clear understanding complex predictive models.

Authors: Przemysław Biecek, PhD[a,c], Marcin Chlebus, PhD[b], Janusz Gajda, PhD[b], Alicja Gosiewska[a], Anna Kozak[a], Dominik Ogonowski[d], Jakub Sztachelski[c], Piotr Wojewnik, PhD[e].

Institutions:

a. Faculty of Mathematics and Information Science, Warsaw University of Technology, ul. Koszykowa 7, 00-662 Warsaw.
b. Faculty of Economic Sciences, University of Warsaw, ul. Długa 44/50, 00-241 Warsaw.
c. Faculty of Mathematics, Informatics, and Mechanics, University of Warsaw, ul. Banacha 2, 02-097 Warsaw.
d. Data Juice Lab sp. z o.o., ul. Platanowa 11, 03-054 Warsaw.
e. Polish Credit Bureau S. A. (Biuro Informacji Kredytowej S.A.), ul. Zygmunta Modzelewskiego 77a, 02-679 Warsaw.

Corresponding author:

Marcin Chlebus, PhD
Faculty of Economic Sciences, University of Warsaw, ul. Długa 44/50, 00-241 Warsaw
mob: +48501305291; e-mail: mchlebus@wne.uw.edu.pl

Funding: The research was partially financed by the Polish Credit Bureau S.A. as internal B+R project.

Abstract

Rapid development of advanced modelling techniques gives an opportunity to develop tools that are more and more accurate. However as usually, everything comes with a price and in this case, the price to pay is to loose interpretability of a model while gaining on its accuracy and precision. For managers to control and effectively manage credit risk and for regulators to be convinced with model quality the price to pay is too high. So, it prevents them from using advanced models due to the lack of their interpretability. In this paper, we show how to take credit scoring analytics in to the next level, namely we present comparison of various predictive models (logistic regression, logistic regression with weight of evidence transformations and modern artificial intelligence algorithms) and show that advanced tree based models give best results in prediction of client default. What is even more important and valuable we also show how to boost advanced models using techniques which allow to interpret them and made them more accessible for credit risk practitioners, resolving the crucial obstacle in widespread deployment of more complex, "black box" models like random forests, gradient boosted or extreme gradient boosted trees. All this will be shown on the large dataset obtained from the Polish Credit Bureau to which all the banks and most of the lending companies in the country do report the credit files. This huge extent of data ensures high quality of the model inputs and objectivity of the conclusions. In this paper the data from lending companies were used. The paper then compares state of the art best practices in credit risk modelling with new advanced modern statistical tools boosted by the latest developments in the field of interpretability and explainability of artificial intelligence algorithms. We believe that this is a valuable contribution when it comes to presentation of different modelling tools but what is even more important it is showing which methods might be used to get insight and understanding of AI methods in credit risk context.

Enabling Machine Learning Algorithms for Credit Scoring - Explanatory Artificial Intelligence (XAI) methods for clear understanding complex predictive models.

**1. Introduction**

Credit Issuance and Collection is at the heart of banks' and lending companies' activities. Lending generates most of their profit (due to the "Zero or Negative Interest Rate\" environment) but in the same time, it may cause severe losses, hurting the bottom line. Thus forecasting of Probability of Default "PD" of a borrower is clearly an important issue. It is now clear that the idea of credit scoring appeared a long time ago. Almost 70 years ago, the foundations of what we call today credit scoring were given [1]. Originally, scoring has only been used for individual customers. Since then institutions have begun to store and collect information from credit applicants to be able to better asses their creditworthiness. Furthermore, the Scoring techniques were extended to create Corporate Rating Models deployed for Small and Medium Enterprises ("SME") and Large Corporations. We can then state [2] that credit scoring involves developing rules and decisions based on expert knowledge and numerical methods which allow for effective scoring decisions. Especially, in modern world when the number of credit products offered is increasing the ability to precisely differentiate among bad and good clients not only reduces the losses of the company, but also increases its competitiveness in the market. Even 1% of increase in accuracy of predictions of borrower willingness to pay, may significantly improve the profit of financial institutions [3].

The need for accurate forecasting in finance and in various other fields caused rapid development of statistical models [4]. Multitude of methods cause that many problems are nowadays not only modellable, but also can be approached from different angles thus we are given an unprecedented possibility to get a deep insight in the nature of considered phenomenon [6]. However such, one might say comfortable solution comes with drawbacks. Modern models and statistical techniques involving complicated mathematical and statistical transformations, advanced numerical routines and demanding algorithmic approach are more and more accurate but in the same time more complex and thus much harder to interpret. As it is stated in [5] during the variable selection phase not only criterion of statistical significance determines the model development, but also the interpretability of the variables and model architecture [5]. It is also highlighted in Basel Accords of Basel Committee on Banking Supervision and local regulations like European Central Bank regulations for European Union. One should mention here General Data Protection Regulation (GDPR) especially articles 13,14 and 22 [7] and the important White Paper on Artificial Intelligence [8]. In this paper we would like to use the unique opportunity provided by the Polish Credit Bureau (40.1mio accounts, for 24mio individuals and 1.1mio SME companies), by supplying us with reliable, objective and large dataset. The dataset combines information about clients from most on-line lending companies in Poland. It is worth to mention here that since lending companies are not limited by the Basel requirements when it comes to use modern machine learning (ML) algorithms, then this work is a valuable contribution as a extensive comparison of the best practices in credit scoring like logistic regression, logistic regression with weight of evidence (WOE) transformation with recent developments in ML like extremely boosted random trees. Such comparison although not new in the literature always provides additional evidence, and in the case considered in this article, namely lack of limitations due to Basel laws, the methods proposed here can be actually applied to solve real problems. Moreover our approach applied in this article is to compare methods commonly used in credit risk modelling with the new ML approach applicable in this context but also was proven to be successful in solving other types of problems  during Kaggle modelling competitions.

    Next we would like to face the "black-box" reputation of modern statistical algorithms and present possible tools one can use to make them explainable. We clarify certain ML techniques. by showing the significance of the factors affecting the results and performing so

called "what if" analysis. In this way we complete the last, one might say the most important, piece of the problem of credit risk modelling with modern AI tools - using some easy to understand information we do explain how the machine learning predictions were made.

The rest of the paper is organised as follows. Section 2 discusses previous work in the subjects of scorecard modelling and model explanations. Section 3 discusses various algorithms for scorecard construction ranging from the most classical ones to most novel. In Section 4 we discuss and analyze experimental results. We deeply discuss also how the new methods for complex model explanations can help with the understanding of modern machine learning tools. Section 5 concludes the paper.

## 2. Related work

Classical credit scoring involves building a model which estimates the probability of an applicant not defaulting within some specified period, usually one year. (as recommended in CRD IV/Basel). The usual approach is to apply the statistical techniques (Logistic Regression, Linear Discriminant Analysis, Decision Trees) to describe the behaviour of applicants and use it to assign some scores to the applicants. The scores should reflect whether the applicants are expected to perform good or bad [6]. Such an approach is widely used across risk management departments in banks, insurance companies or other financial institutions. The best practices involves also appropriate features engineering (Weight of Evidence transformation) and feature selections methods (representatives of collinear variables, general to specific or stepwise selection [6]) .

Recent development in the field of statistical modelling encourages to use more and more advanced techniques. One can build a quite comprehensive list of research articles exploring possibility of application of modern techniques in predictive analytics in the credit industry. The fundamental yet powerful techniques involve usage of logistic regression [6,8] and decision trees [9,10]. The latter and the former are the standard tools, however one can list also probit regression [11], random forests [12], k-nearest neighbours (KNN) [13-14], artificial neural networks (ANNs) [15-16], support vector machines [17] and many many others. In [18] authors compare 41 various novel algorithms for classification, moreover they also discuss performance measures and techniques that allow to reliably compare different classifiers. In [19] authors present systematic literature review related to theory and application of classification techniques for credit scoring.

      Since even a tiny improvement leads to noteworthy future savings one cannot underestimate the importance of proper model. As modern statistical techniques have proven its usage in the information science and capability to classify a customer as a good or bad payer, in the context of credit risk, with respect to its characteristics, it is understandable to expect the future of credit risk modelling among those methods. However what might be considered as biggest strength of these approach has some important drawback when one tries to apply them in business applications, where each decision has to be meaningful and the user has to understand the underlying reasons why the model assigns the customer to be a defaulter. This statement is clearly visible in new Basel Accords which are standard guides for financial companies. To be complainant with those regulations one needs to search for tools and concepts that not only allow for accurate predictions but in the same time allow models to be more understandable. As it is stated in [20] eXplainable Artificial Intelligence (XAI) allows complex type black-box models to be more understandable and thanks to XAI methods one is able both to detect the most important features, but also to perform the what-if analysis and get more insight in the decision process of an algorithm [21]. Most of the XAI methods possess also the advantage of being the so called model agnostic, thus one can apply the same techniques over a range of complicated ML models. This allows for simple and universal comparisons of various models. In recent years we observe significant increase in development and

applications of XAI methods, see LIME [22], SHAP [23], DALEX [21] or InterpretML [24] – the 2 latter are a whole XAI frameworks.

Among various methods and tools there are couple deserving some more attention. Thus the simplest method is the permutation feature importance (PFI) [25]. The method aims to assess the validity or importance of each feature. First we need to choose the accuracy metric to be applied. we will be investigating which in In the case of credit risk classification models it will be our case was the area under receiver operating characteristics curve (AUC). Then we analyse the importance of each feature by calculation the changes of AUC when we run model with permuted values of that feature. The other method of feature assessment is the partial dependence plot (PDP) [26]. PDP shows the marginal influence of one or two features on the forecast outcome of certain machine learning algorithm. The dependency can be of linear, monotonic or some more complex form. PDP is calculated by specifying a specific value of one or two input variables and then averaging results of forecasts for all possible values of remaining explanatory variables. The PDP is helping to understand model on the global level.

Another important method are the so called Break Down plots (BD) which decompose the specific model prediction (instance level) by explanatory variables to show the impact of each feature on the final outcome. BD allow the user to receive a clear summary of effect that each variable has on expected model forecast. To perform a simple what-if analysis, namely to estimate how model response will change following the changes of the input variables, one can also use Ceteris Paribus plots (CP) [27,28]. This straightforward idea gives the user a first impression how some change of an input variable influences the forecast assuming other variables to be constant. One can observe that PDP is an average of many CP and in that context this method can be perceived as a global compared to local interpretation via CP.

**3. Algorithms for scorecard construction**

In present paper we used the approach proposed in [18] which compares various models for prediction whether the borrower will default or not. This is classical classification problem [29] which can be tackled by various ways. Our data can be represented as $D = (y_i, x_i)_{i=1}^n$, where $y_i$ is explanatory variable taking values 0 or 1 for good and bad borrower respectively and vector $x_i$ represents values of explanatory variables. We picked some algorithms presented in [18] but also supplemented our study with recently developed models. Thus we compare here application of well-known in the industry logistic regression [6] which allows for modelling of the likelihood of default of a borrower. We also propose to study logistic regression with weight of evidence (WOE) transformation approach [6]. The WOE method allows capturing some nonlinear dependence between transformed predictions and the explanatory variable. The other methods are modern machine learning algorithms. Here we applied random forest approach [30] based on construction of many decision trees. While each decision tree produces some predictions, thus the final outcome of the whole random forest has to be chosen among them, and the choice is based on the frequency of particular values appearance (so called majority voting). Such grouping of many weak classifiers (bagging) has a plausible property in reduction of the forecast variance [31].

The last approach involves usage of recently developed boosting methods: gradient boosting (GBM) and extreme gradient boosting (XGB) models [32-33]. On the contrary to the bagging approach we don't build and group all the classifiers at once. In the boosting approach the classifiers are built one after another and they are targeted to improve on the misclassified cases from the previously prepared model. Similarly to the bagging case also in the boosting approach the weak learners will use the classification trees.

Application of the modern methods in the context of credit risk modelling and their comparison with well known, accepted and commonly applied methods is not the single aim of this paper. What might be even more important we would like to discuss how XAI tools can be used for these models and how they can help us to understand the logic behind the decisions of complex algorithms, and how particular feature influences the forecast. Since the XAI methods explored in this paper are "model-agnostic" (i.e. can be used to get more insight in any model) we also check how they do perform in the case of classical methods. Such insight has a valuable advantage since one may validate the XAI approach directly to classical approach of inspecting the sign and magnitude of the model coefficients.

**4. Experimental results**

Dataset we use in this paper comes from Polish Credit Bureau, it consists of over 5 million observations for lending companies' clients with 1 729 variables and covers the time span from 1.10.2017 up to 30.05.2019. Due to the assumption of 6 month default window we have limited the modelling sample to cover the period till 30.11.2018. According to company policy data were fully anonymised. In order to assess the quality of the analysed models dataset was split to 4 sets, namely training set, on which all mentioned models were fitted, test set which was used to assess the accuracy and stability of the models, out of sample set covering period 1.10.2017-30.08.2018 which did not take any role in procedure of fitting models but were later used to assess their efficiency. We also considered out of time set for observations of the last three months in period 31.08.2018 a 30.11.2018 on which we tested how well the model generalizes to new, previously not observed data and whether its predictions are stable in time. This last issue has significant importance in the context of assessment how well model reacts in rapidly evolving environment and whether it absorbs the changes.

Before the modelling part, all variables were examined in the context of descriptive statistics, missing data were filled by means to avoid disruptions of linear models and for some variables we used the so called dummy enconding. Since the number of predictors was high (i.e. 1 729) which is obviously not a problem for modern machine learning techniques but for classical methods like logistic regression it might pose some problems we decided to use some preselection method to find the most appropriate variables. Our method based on xgboost approach, the variables were divided into subsets, which were distinguished by the number of unique values. The division threshold was set to 300 which was also the median of the number of unique values of the variables. On such sets, xgboost models were built, indicating predictive power of 81 variables. Then for the selected variables the Kolmogorov- Smirnov (K-S) statistic was calculated and the values below 0.1 pointed variables to be removed, what ultimately led to the selection of 63 variables. Based on that set we have estimated our models.

What we would like also to emphasize is that we tested also other models, such as neural networks, naive Bayes, regularised logistic regression, discriminatory analysis., as well as, random forest, gbm and xgboost trained on more conservatively selected 34 variables. The hyperparameters were tuned based on random search approach. To assess all the models quality we used Gini measure and rejection was based on the expert knowledge and Gini value below 0.6.

In the Figure 1 we present results of our models on the training and test sets. As one can see the results of all the models are quite good. The random forest ( rf_63_RS) has an extremely high Gini value (1.0) on the training, whereas it is the effect of excessive depth of trees in random forest, and actually - was expected. The model achieved also the highest Gini value (0.76) on the test set. GBM (gbm_63) resulted also in the very high scores (0.68) on both sets. Logistic regression with WOE transformation (reg_log_63_woe) model is the best model in the logistic model class (pure logistic regression model denoted here by reg_log_63), and it scores on the test set even better (0.65) than the xgboost (xgb_63_RS) model (0.64), but visibly weaker than the gbm_63 and rf_63_RS models. All the other models achieved worse results and therefore are not discussed further.

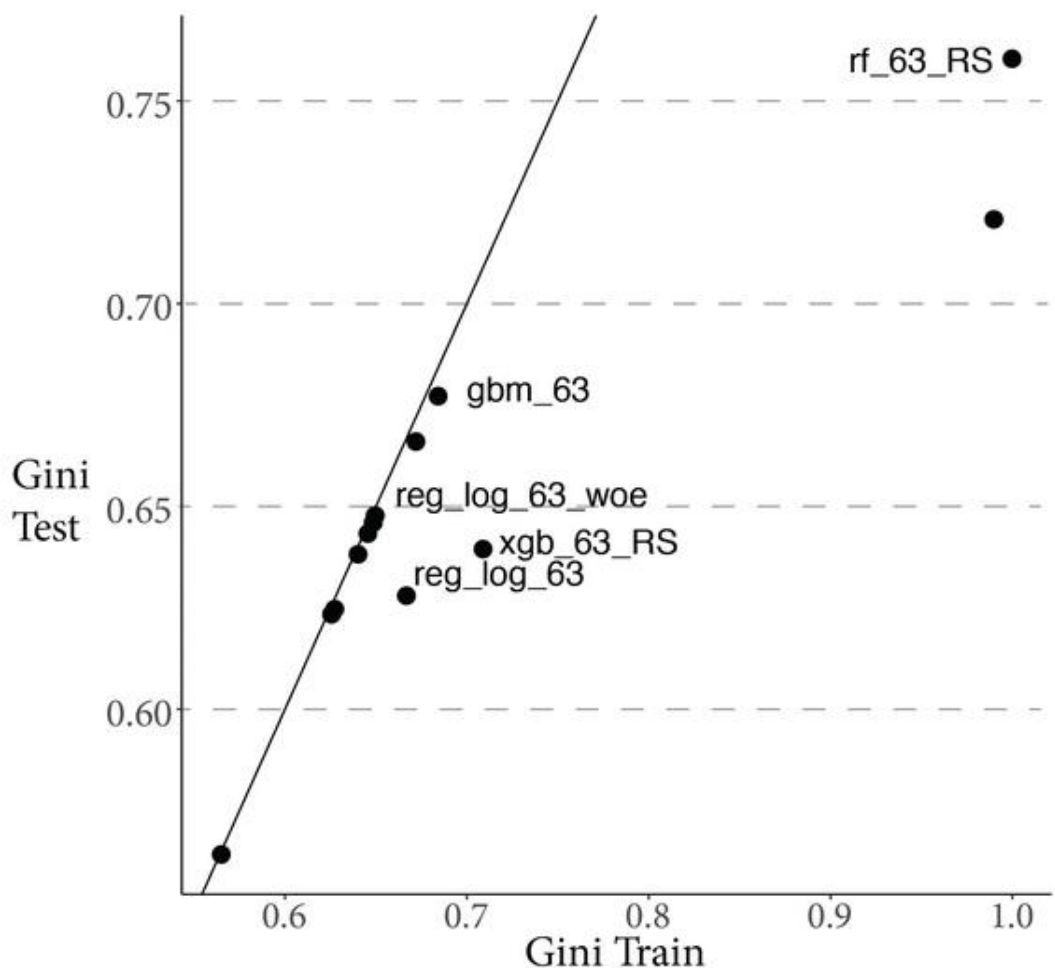

Figure 1: In the graph, the results for 5 key models are marked with their names: Gradient Boosting Machine (gbm_63), Random Forests (rf_63_RS), logistic regression with and without WOE transformation (reg_log_63_woe, reg_log_63) and xgboost model (xgb_63_RS)

The results of the model comparison for the out-of-sample data set are shown in Fig 2. One can easily see that Gini's values on the out-of-sample (green points) for all models are comparable to the results on the test sample (black points). The difference between Gini on training, test and out-of-sample is at the level of 0.001 except for random forest algorithms due to their specificity.One can also notice in Fig 2 a decrease in Gini values on the out-of-time test sample (1% - 9%) (the decrease for individual models is marked with a vertical line and red points). The biggest decrease concerns the best models - random forests; the RF 63 model lost about 10 p.p.. The best model on the out-of-time sample is GBM 63. The decrease for this model on the out-of-time sample is much smaller than for the random forest (only 1%). It also has higher results than logistic regression models (by ~3 p.p. of Gini value).

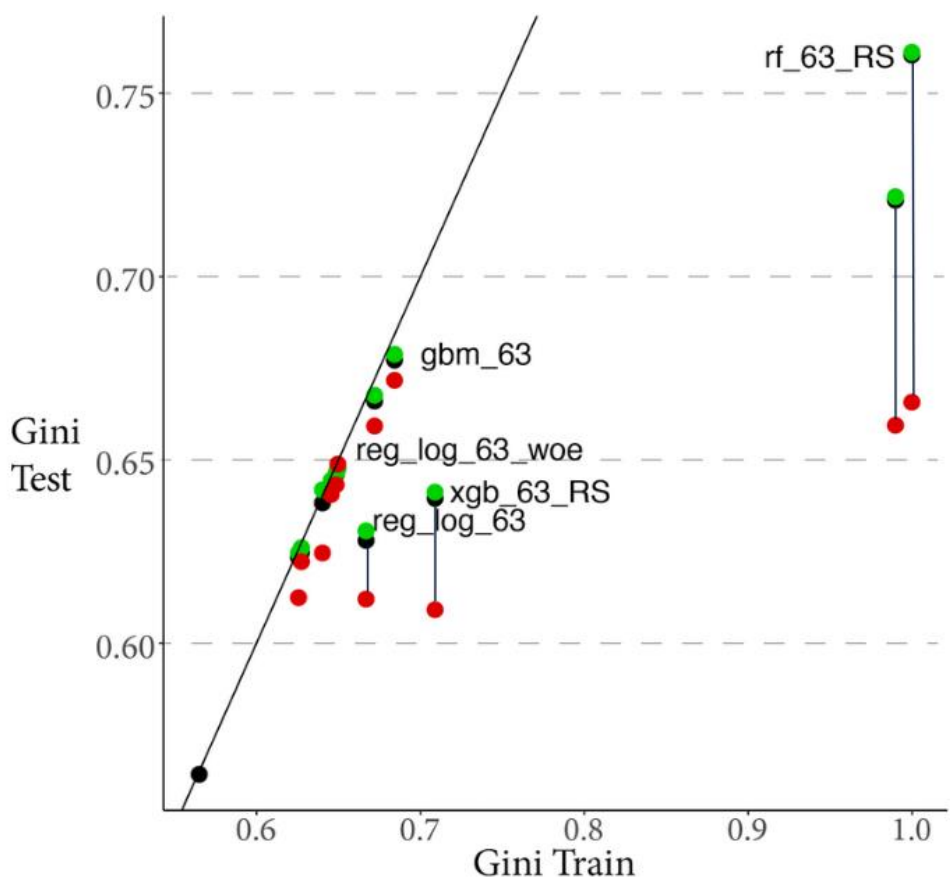

Figure 2: In the graph, the results for 5 key models are marked with their names: Gradient Boosting Machine (gbm_63), Random Forests (rf_63_RS), logistic regression with and without WOE transformation (reg_log_63_woe, reg_log_63) and xgboost model (xgb_63_RS). Out of sample Gini values are denoted by green dots. Black dots denote the results on the test sample while red points denote the out-of-time results. One sees the clear fall in the Gini values for all models when calculated on the out-of-time sample.

| Name of the model | Gini Train | Gini Test | Gni out of sample | Gini out of time | K-S out of time | Learning time (min) | Prediction time (sec) | Comment |
|---|---|---|---|---|---|---|---|---|
| rf_63_RS | 1 | 0,76 | 0,76 | 0,67 | 0,51 | 40 | 7 | Time for predction for 1-1000 obs the same |
| gbm_63 | 0,68 | 0,68 | 0,68 | 0,67 | 0,52 | 30 | 0,01 | Time for predction for 1-1000 obs the same |
| reg_log_63_woe | 0,65 | 0,65 | 65 | 0,64 | 0,49 | 0,17 | <0,01 | |
| reg_log_63 | 0,65 | ,65 | 0,65 | 0,6 | 0,49 | 1,50 | <0,01 | |
| xgb_63_RS | 0,71 | 0,64 | 0,64 | 0,61 | 0,46 | 0,20 | 0,01 | |

Table 1. Detailed model comparison. Here we provide Gini measures on each of the analysed sets, K-S statistics on the out-of-time set, learning and prediction time of for all the models considered in the paper. One can see that the Gradient Boosting Machine (gbm_63) gives overall the best results, especially when analysing its stability across all sets.

The detailed results of the model comparison are presented in Table 1. It can be seen that for the random forest model, the Gini value decreases with each subsequent set, but is still at the highest level compared to the other models. Additionally, it is worth noting the gradient boosting machine (GBM), its results are stable on all sets (0.67 - 0.68 Gini). The reference model, i.e. logistic regression with WOE transformation, also achieves stable results (0.64 - 0.65 Gini), but it is 3 p.p. lower than the recommended model. The efficiency measured by the Gini metric is consistent with the results of Kolomogorov-Smirnov's statistics, i.e. models with a larger Gini also have higher statistics. The recommended model, i.e. GBM with 63 variables, has not only the highest value of the Gini measure, but also K-S statistics. Analyzing the results presented in the table, it is worth noting that the prediction time for random forests is fundamentally higher (7-11 seconds) than for GBM class models (<0.01 seconds).

Explainability

After analysis of the algorithms performance we would like to tackle the another objective of this paper, namely to investigate the tools available to gain some understanding of the models in the context of global and local interpretation. Since as we mentioned variable names in the analysed data set were anonymised due to safety reasons – variable names are not self-explanatory, however still we will be able to see results and discuss properties and the outcome of the method.

Global interpretations

By using the global interpretation methods one can assess the general role of individual variables in the model and their average impact on predictions. The tools we can apply here are PDP profiles discussed earlier and the universal method based on permutations, the so called permutation feature importance (PFI). Let us start with the latter method, where the concept is straightforward and easy to use. One can clearly see that this method is universal and can be applied to any considered model. In Fig 3 we present the results in the context of GBM model. What can be inferred here is that the biggest drop in AUC by 0.22 we observe for the feature coded as AGR_417. Thus in the other words, after randomly shuffling this feature many times and use in the model we observe the highest mean decrease in accuracy of the model. Therefore this variable is the most influential one in discrimination of defaulters and non-defaulters. One can say that the proper assessment of this feature accounts in almost 7% in the process of decision making. Second important variable is OAG_7 which accounts in almost 3% in process of discrimination between defaulters and non-defaulters. In Fig 4. we present PFI for gradient boosting model, here also one can assess the importance of the feature OAG_7 which for this model accounts for almost 2% when making decisions on differentiation between defaulters and non-defaulters. What is important here that PFI can be used for different models and by investigating it for various cases we can build our intuition which variables are the most important in the process of discrimination. Here we observe that in both approaches out of 63 variables, maybe in different order however by looking at the scale on X axis differences are not very high, both models selected as the most important almost the same variables, thus those are most substantial and their perturbation by random shuffling brings the highest error in the accuracy of various models. What is more, it can be easily seen that a role of each variable is not fundamentally higher than the others and thus a risk of too big concertation of the model for a one (several) variable(s) is not crucial. This is very often expected by the Credit Risk Management managers and decision makers, as it's considered as more stable solution.

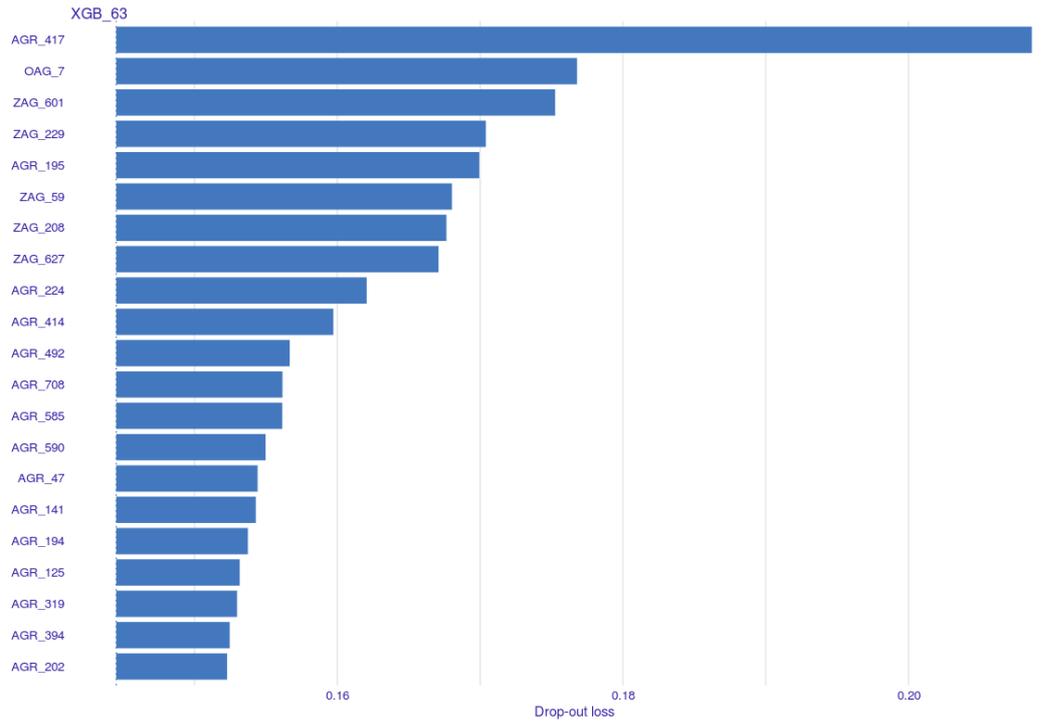

Figure 3: Permutation Feature Importance results obtained with XGB classifier based on 63 variables (XGB_63). The X axis depicts AUC decrease after permutation of selected variable. The bigger the difference, the more important variable is. Big differences mean that the model makes worse predictions without "knowing" the values of the selected variable.

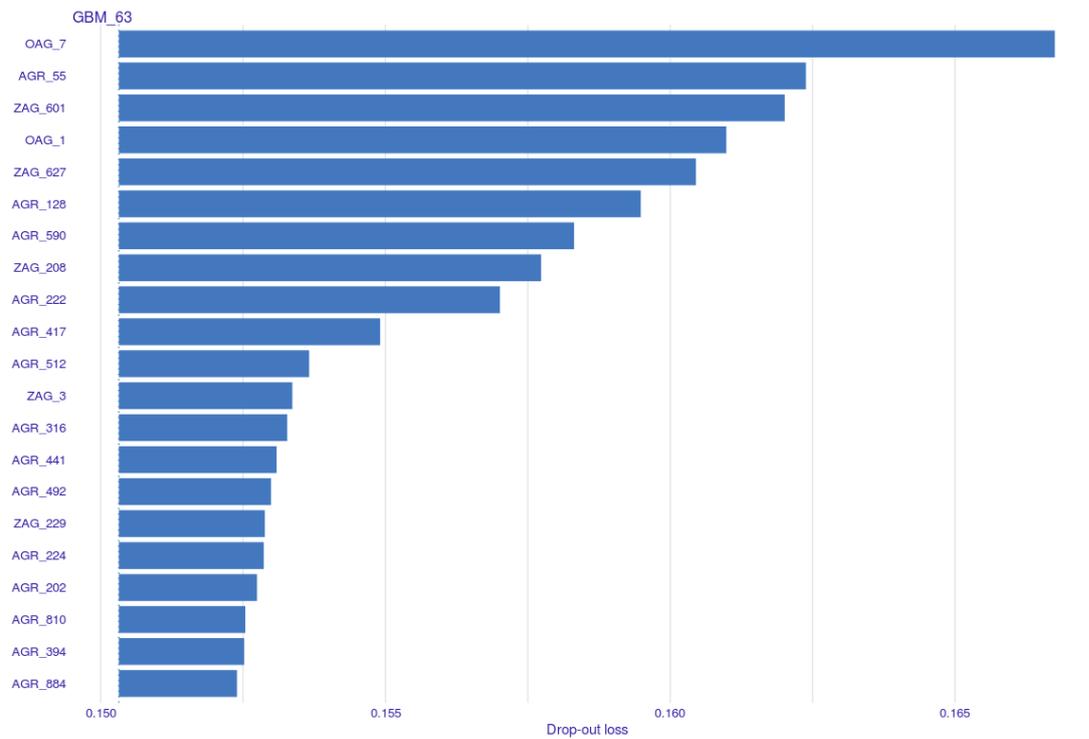

Figure 4: Permutation Feature Importance results obtained with GBM classifier based on 63 variables (GBM_63). The X axis depicts AUC decrease after permutation of selected variable. The bigger the difference, the more important variable is. Big differences mean that the model makes worse predictions without "knowing" the values of selected variable.

Basing on the Fig 5 and PDP plot we would like to get some more insight on the most important feature – identified with PFI method. The PDP profile shows the dependency between the values of analyzed variable and average prediction of given model. From the Fig 5 one can assess that all models predictions behave similarly with respect to variable OAG_7, namely average predictions decrease with respect to higher values of that feature. The similar behaviour is observed both for classical (WOE and GLM) and modern (GBM, XGB) models, thus with the classical approaches one can validate the reliability and intuitiveness of the ML algorithms. Observing how PDP profiles change for each model we may conclude that all models try to capture the same relation (first faster decline, than stability). Especially GMB and Logistic Regression with WOE transformation have very similar profiles, the only difference is that GBM is more flexible and therefore could adjust better to relation observed in data.

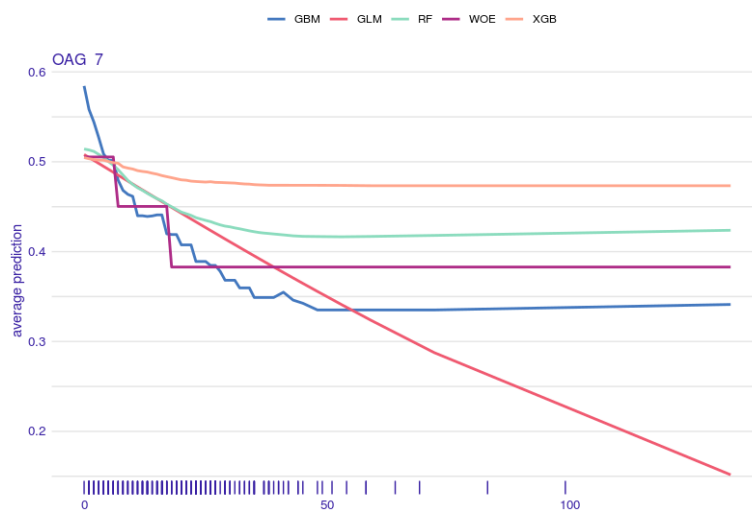

Figure 5: Partial Dependence Plot results for models considered in the paper. All the models describe a similar relationship between the OAG_7 variable and the predictions. Namely we observe lower average prediction values when OAG_7 increases, in other word probability of classification a client as defaulter decreases on average when OAG_7 increase.

Local methods

Local methods aim to explain the crucial factors determining particular prediction. The local equivalent of a PDP profile for a single observation is the CP profile which shows the change of prediction value when changing the value for the analyzed variable keeping all other variables fixed. While in the PDP plot we were averaging predictions for different values of all features except the selected one, thus by averaging many CP profiles we simply obtain a PDP plot. In Fig 6 we present CP plots for selected features for particular customer. The blue dots denote the current values of selected variables. Let us consider the first plot – for OAG_7 variable. While keeping all the other variables constant one observes that increase from the current state will decrease the prediction value i.e. decrease the probability of default, while if we consider smaller values for OAG_7, then the probability of default will increase. Moreover we may see, that CP for a given client has similar shape as PDP for the xgb model, but probabilities vary from 0.12 to 0.02 (in PDP form 0.52 to 0.48), the move towards smaller probabilities is reflecting a smaller risk caused by other variables. For other variables like ZAG_601, ZAG_627 we can observe that if we increase their values we increase PD while shifting values in any direction for AGR_55 and OAG_1 will not influence PD much.

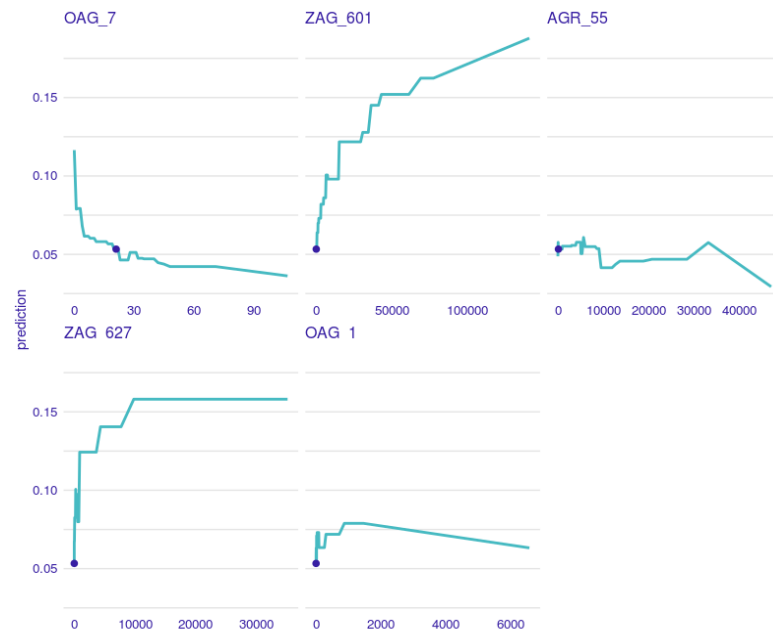

Figure 6. Ceteris Paribus plots for various variables for GBM model. Blue dots denote current values of those features. An increase in OAG_7 will slightly decrease PD for a selected customer, while a decrease in OAG_7 would significantly increase the level of forecast PD. An increase in ZAG_601 and ZAG_627 would significantly increase the customer's PD forecast. The customer is not very sensitive to the variables AGR_55 and OAG_1.

Another useful tool for visualization of predictions are BD plots which decompose predictions for selected customer to assess contribution of each individual variables. In Fig. 7 we present BD plots. Here we can see that final PD (blue bar) is 0.051. The average prediction in the analyzed sample is 0.559 (Intercept only), i.e. the projected PD is on average 56%. Based on BD plot we observe that the OAG_7 variable reduces the PD forecast by 0.073 and the AGR_222 variable by 0.08. Similarly we can read individual contributions for remaining features. In this case, the analysis of the BD plots is limited as variables are anonymised, however, in case the variables would be known, we may also check and confirm business interpretation of a given results.

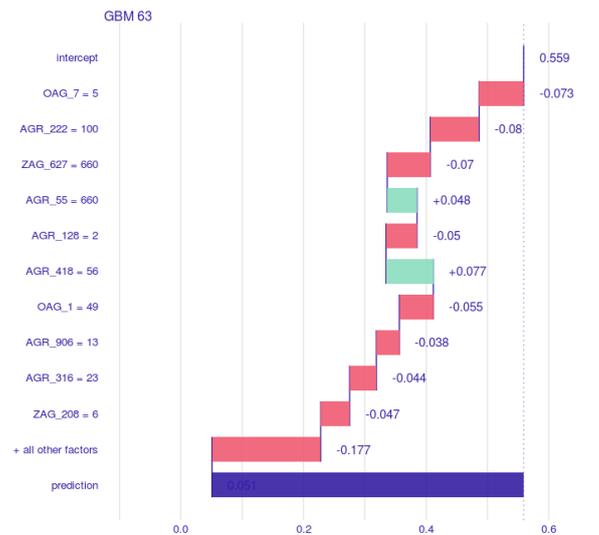

Figure 7. Break Down plots for selected customer. The actual values are presented on the Y axis while on X axis we have individual contributions. Thus our actual prediction (blue bar) is 0.051, and we can infer that the average prediction is 0.559 (Intercept only) while variable OAG_7 reduces the PD forecast by 0.073.

Business Context

The application of the findings in the business context, specifically for the Risk Management is profound. It clearly shows the ability of more advanced modelling techniques to better differentiate risk by providing the stronger models while maintaining their stability. The wider deployment of advanced modelling techniques in Risk Management was limited by the complexity of these model. On the one hand we can observe some internal reasons – to build the highest quality of the service, full transparency of decisions and also the feedback loop from the customers - the model owners and users should poses the ability of explaining to the customers, including individual customers, the reasons for the credit decisions, especially if they are based on automated profiling. On the other hand there are also external reasons. The consumer protection mechanism makes it difficult for financial institutions to simply say no to loan prospects, without any explanation. Not mentioning the business and reputational impact on the rejected customers. The regulations both in the US "Equal Credit Opportunity Act" and the GDPR EU have mandated the need of the lender to explain the decision taken through automated processes. So, the Bank's Communication and User Experience folks had to first understand the underlying reasons for the rejections, before explain them to customers. Up to the moment it has been a difficult and challenging task for the ML models, but it has been accomplished with some simple rules extracted from the models with XAI methodology:

So, Dear Customer,

- please, pay on time as any delay in instalment payment 30 days will significantly decrease of any banking financing in the future,
- please, do notexceed the limit of your revolving facilities,
- please, reduce the usage of the revolving facilities, as the high usage negatively impacts your scoring,
- please, develop your credit history by applying for a card or a loan and servicing it regularly.

The primary Business utilization of our findings is to provide empirical evidence that the advanced modelling techniques like Gradient Boosting provide for better models. The secondary utilization is to give evidence that the advanced models can be interpreted with a set of XAI tools, providing a required regulatory compliance, and thirdly that there are tangible benefits from stronger and stable models, making a business case viable for now "not-so-black-box models".

## 5. Conclusions

In this paper we discussed and compared various methods for modelling of default probability. This, as we presented throughout the paper, is very important problem for financial institutions around the world since it directly affects their profitability. Such comparisons are thus important and one can say that this subject never ends. Each and every subsequent presentation and comparison of new techniques is of separate importance then. However this paper discusses also fundamental issue which always appears when one wants to use advanced machine learning in financial modelling. This is how to explain, and understand complex model.

Such issue is very important for all. For model developers to better understand the models, to get the basic insight which allow to confront model results with intuition and common sense. For executive board to understand how the model works, which features are most important and affect results the most. Lastly for model recipients who want to know why their application for a credit was rejected.

What is also important to stress is, that analysis of model explanations can be useful in better understanding of data itself. Moreover it can be helpful in detection of various data problems which might speed up the analyst work in preparing variables.

We believe that this paper brings up the whole problem of default modelling which are the proper statistical technique but also its meaningful explanation allowing to understand how each and every decision was made.

**References**

[1] Durand, D., (1941). Risk elements in consumer instalment financing, in: National Bureau of Economics, New York.

[2] Thomas, Lyn C., Edelman, David B. and Crook, Jonathan N. (2002) *Credit Scoring and its Applications* , Philadelphia, USA. SIAM, 246pp.

[3] Hand, D.J. and Henley, W.E. (1997) Statistical Classification Methods in Consumer Credit Scoring: A Review. Journal of Royal Statistical Society, 160, 523-541.  https://doi.org/10.1111/j.1467-985X.1997.00078.x

[4] Leo, M., Sharma, S., & Maddulety, K. (2019). Machine Learning in Banking Risk  Management: A Literature Review. *Risks*, *7*(1), 29. MDPI AG. Retrieved from  http://dx.doi.org/10.3390/risks7010029

[5] Martens, D., B. Baesens, T. Van Gestel, and J. Vanthienen. 2007. "Comprehensible Credit Scoring Models Using Rule Extraction from Support Vector Machines." European Journal of Operational Research 183:1466–1476

ENABLING MACHINE LEARNING ALGORITHMS FOR CREDIT SCORING    23
[6] Siddiqi, N. (2006). Credit risk scorecards: Developing and implementing intelligent credit scoring. Hoboken, N.J: Wiley.

[7] https://gdpr-info.eu/

[8] https://ec.europa.eu/info/publications/white-paper-artificial-intelligence-european-approach-excellence-and-trust_en

[8] Campbell, J.Y., Hilscher, J., & Szilagyi, J. (2008). In search of distress risk. Journal of Finance 63 (6), 2899–2939. doi:10.1111/j.1540-6261.2008.01416.x

[9] Frydman, H., E. I. Altman and D-L. Kao. (1985): Introducing recursive partitioningfor financial classification: the case of financial distress. Journal of Finance (March): 269-291. https://doi.org/10.1111/j.1540-6261.1985.tb04949.x

[10] Davis, R. H., Edelman, D. B., & Gammerman, A. J. (1992). Machine-learning algorithms for credit-card applications. *IMA Journal of Management Mathematics*, *4*(1), 43-51. https://doi.org/10.1093/imaman/4.1.43

[11] Tsaih, R., Liu, Y. J., Liu, W., & Lien. Y. L. (2004). Credit scoring system for small business loans. *Decision Support Systems 38(1),* 91–99. https://doi.org/10.101/S0167–9236(03)00079–4

[12] Kruppa, J., Schwarz, A., Arminger, G., & Ziegler, A. (2013). Consumer credit risk: Individual probability estimates using machine learning. Expert Syst. Appl., 40, 5125-5131.

 [13] Henley, W., & Hand, D. (1996). A $k$-Nearest-Neighbour Classifier for Assessing Consumer Credit Risk. Journal of the Royal Statistical Society. Series D (TheStatistician), 45(1), 77-95. doi:10.2307/2348414

 [14] Brown,  I.,  Mues,  C. (2012). An experimental comparison of classification algorithms for imbalanced credit scoring data sets. Expert Systems with Applications 39 (3), 3446–3453. https://doi.org/10.1016/j.eswa.2011.09.033

[15] Desai, V.S., Crook, J.N, Overstreet, G.A. (1996). A comparison of neutral networks and linear scoring models in the credit union environment. European Journal of Operational Research 95, 24-37. https://doi.org/10.1016/0377-2217(95)00246-4

[16] Malhotra, R.; Malhotra, D. K. (2002). Differentiating between good credits and bad credits using neuro-fuzzy systems. European Journal of Operational Research, 136(1), 190-211 https://doi.org/10.1016/S0377-2217(01)00052-2

[17] Vapnik, V.N. (1995) The Nature of Statistical Learning Theory. Springer, New York. http://dx.doi.org/10.1007/978-1-4757-2440-0

[18] Lessmann, S., Baesens, B., Seow, H. V., and Thomas, L. C. (2015). Benchmarking State-of-the-Art Classification Algorithms forCredit Scoring: An Update of Research. European Journal of Operational Research (247:1), pp. 124-136. https://doi.org/10.1016/j.ejor.2015.05.030

[19] Louzada, F., Ara, A., & Fernandes, G. (2016). Classification methods applied to credit scoring: A systematic review and overall comparison. Surveys in Operations Research and Management Science2016 21:117–134. https://doi.org/10.1016/j.sorms.2016.10.001

[20] L.H., Bau, D., Yuan, B.Z., Bajwa, A., Specter, M., Kagal, L. (2018). Explaining explanations: an approach to evaluating interpretability of machine learning. arXiv preprint. arXiv:1806.00069.

[21] Biecek, P. (2018). DALEX: Explainers for Complex Predictive Models in R. Journal of Machine Learning Research, 19(84), 1-5. http://jmlr.org/papers/v19/18-416.html.

[22] Ribeiro, M.T., Singh, S., & Guestrin, C. (2016). Why should I trust you? Explaining the predictions of any classifier. 2016

ArXiv https://arxiv.org/abs/1602.04938.

[23] Lundberg S.M., & Lee, S.-I. (2017). A unified approach to interpreting model predictions. In Proceedings of the 31st International Conference on Neural Information Processing Systems (NIPS'17) (pp. 4768–4777). Curran Associates Inc., Red Hook, NY, USA.

[24] Nori, H., Jenkins, S., Koch, P., & Caruana, R. (2019). InterpretML: A Unified Framework for Machine Learning Interpretability. ArXiv, abs/1909.09223.

[25] Fisher, A., Rudin, C., & Dominici, F. (2018). Model class reliance: Variable Importance measures for any machine learning model class, from The "Rashomon" perspective. arXiv preprint arXiv:1801014892018.

[26] Friedman, J.H. (2001). Greedy function approximation: A gradient boosting machine. Ann. Statist. 29(5), 1189--1232. doi:10.1214/aos/1013203451.

[27] Biecek, P., Baniecki, H.,Izdebski, A., & Pekala, K. (2019). ingredients: Effects and Importances of Model Ingredients.

[28] Goldstein, A., Kapelner, A., Bleich, J., & Pitkin, E. (2015). Peeking Inside the Black Box: Visualizing Statistical Learning With Plots of Individual Conditional Expectation. Journal of Computational and Graphical Statistics, 24(1), 44–65. doi:10.1080/10618600.2014.907095

[29] Dobson, A. J. (1990) An Introduction to Generalized Linear Models.London: Chapman and Hall.

[30] Breiman, L. (2001). Random Forests. Machine Learning 45, 5–32. https://doi.org/10.1023

[31] Hastie, T., Tibshirani, R., & Friedman, J. H. (2009). The elements of statistical learning: data mining, inference, and prediction. 2nd ed. New York: Springer.

[32] Greenwell, B., Boehmke, B., & Cunningham, J., and GBM Developers (2019).

gbm: Generalized Boosted Regression Models.

[33] Chen, T., & Guestrin, C. (2016). XGBoost: a scalable tree boosting system.arXiv:1603.02754v1